\documentclass[letterpaper]{article} 
\usepackage{aaai2026}  
\usepackage{times}  
\usepackage{helvet}  
\usepackage{courier}  
\usepackage[hyphens]{url}  
\usepackage{graphicx} 
\usepackage{algorithm}
\usepackage{natbib}  
\usepackage{caption} 
\usepackage{comment}
\usepackage{multirow}
\usepackage{comment}
\usepackage{verbatim} 

\usepackage{tikz}
\usepackage{pgfplots}
\usepackage{pgfplotstable}
\usepackage{graphicx}
\usepackage{multirow}
\usepackage{pifont}
\usepackage[export]{adjustbox}
\usepackage{pgf-pie}
\usepackage{tkz-graph}
\usepackage{graphicx}
\usepackage{subcaption}
\usepackage[skip=2pt]{caption}
\usepackage{microtype}
\usepackage{algpseudocode}
\usepackage{xcolor} 
\usepackage{amsmath}
\usepackage{amsfonts}

\definecolor{SlateBlue}{RGB}{106, 90, 205} 
\definecolor{CadetBlue}{RGB}{95, 158, 160} 
\definecolor{BurntOrange}{RGB}{204, 85, 0} 
\definecolor{Goldenrod}{RGB}{218, 165, 32} 

\usepackage{booktabs}

\newcommand*\circled[1]{\tikz[baseline=(char.base)]{
            \node[shape=circle,draw,inner sep=1pt] (char) {#1};}}

\usetikzlibrary{pgfplots.groupplots}
\usetikzlibrary{shadows,patterns,shapes,arrows,decorations.pathmorphing,backgrounds,positioning,fit,plotmarks,calc,spy,matrix}
\pgfplotsset{compat=1.11,
    /pgfplots/ybar legend/.style={
    /pgfplots/legend image code/.code={%
       \draw[##1,/tikz/.cd,yshift=-0.25em]
        (0cm,0cm) rectangle (3pt,0.8em);},
   },
}
\usetikzlibrary{decorations.text}
\definecolor{cadetblue}{rgb}{0.37, 0.62, 0.63}
\definecolor{airforceblue}{rgb}{0.36, 0.54, 0.66}
\definecolor{caribbeangreen}{rgb}{0.0, 0.8, 0.6}
\definecolor{carolinablue}{rgb}{0.6, 0.73, 0.89}
\definecolor{darkgoldenrod}{rgb}{0.72, 0.53, 0.04}
\definecolor{debianred}{rgb}{0.84, 0.04, 0.33}
\definecolor{fuzzywuzzy}{rgb}{0.8, 0.4, 0.4}
\definecolor{grullo}{rgb}{0.66, 0.6, 0.53}
\definecolor{ceil}{rgb}{0.57, 0.63, 0.81}
\definecolor{candypink}{rgb}{0.89, 0.44, 0.48}
\definecolor{calpolypomonagreen}{rgb}{0.12, 0.3, 0.17}
\definecolor{burntsienna}{rgb}{0.91, 0.45, 0.32}
\definecolor{atomictangerine}{rgb}{1.0, 0.6, 0.4}
\definecolor{goldenrod}{rgb}{0.85, 0.65, 0.13}
\definecolor{gamboge}{rgb}{0.89, 0.61, 0.06}
\definecolor{amber}{rgb}{1.0, 0.75, 0.0}
\definecolor{battleshipgrey}{rgb}{0.52, 0.52, 0.51}
\definecolor{darkcerulean}{rgb}{0.03, 0.27, 0.49}
\definecolor{fuzzywuzzy}{rgb}{0.8, 0.4, 0.4}
\definecolor{mediumseagreen}{rgb}{0.24, 0.7, 0.44}
\definecolor{antiquebrass}{rgb}{0.8, 0.58, 0.46}
\definecolor{apricot}{rgb}{0.98, 0.81, 0.69}
\definecolor{asparagus}{rgb}{0.53, 0.66, 0.42}
\definecolor{bananamania}{rgb}{0.98, 0.91, 0.71}
\definecolor{cadmiumgreen}{rgb}{0.0, 0.42, 0.24}
\definecolor{chocolate}{rgb}{0.48, 0.25, 0.0}
\definecolor{cinereous}{rgb}{0.6, 0.51, 0.48}
\definecolor{aliceblue}{rgb}{0.94, 0.97, 1.0}
\definecolor{beaublue}{rgb}{0.74, 0.83, 0.9}
\definecolor{blizzardblue}{rgb}{0.67, 0.9, 0.93}
\definecolor{bittersweet}{rgb}{1.0, 0.44, 0.37}
\definecolor{camouflagegreen}{rgb}{0.47, 0.53, 0.42}
\definecolor{darkolivegreen}{rgb}{0.33, 0.42, 0.18}
\definecolor{darkpastelblue}{rgb}{0.47, 0.62, 0.8}
\definecolor{desertsand}{rgb}{0.93, 0.79, 0.69}
\definecolor{deeppeach}{rgb}{1.0, 0.8, 0.64}
\definecolor{indianred}{rgb}{0.8, 0.36, 0.36}
\definecolor{oldmauve}{rgb}{0.4, 0.19, 0.28}
\definecolor{lightblue}{rgb}{0.68, 0.85, 0.9}
\definecolor{lightcyan}{rgb}{0.88, 1.0, 1.0}
\definecolor{viridian}{rgb}{0.25, 0.51, 0.43}
\definecolor{slategray}{rgb}{0.44, 0.5, 0.56}
\definecolor{manatee}{rgb}{0.59, 0.6, 0.67}
\definecolor{darkbrown}{rgb}{0.4, 0.26, 0.13}
\definecolor{almond}{rgb}{0.94, 0.87, 0.8}
\def\BibTeX{{\rm B\kern-.05em{\sc i\kern-.025em b}\kern-.08em
    T\kern-.1667em\lower.7ex\hbox{E}\kern-.125emX}}
\usepackage{colortbl}
\usepackage{tabulary}

\pgfkeys{%
/piechartthreed/.cd,
scale/.code                =  {\def\piechartthreedscale{#1}},
mix color/.code            =  {\def\piechartthreedmixcolor{#1}},
background color/.code     =  {\def\piechartthreedbackcolor{#1}},
name/.code                 =  {\def\piechartthreedname{#1}}}

\usepackage{newfloat}
\usepackage{listings}
\DeclareCaptionStyle{ruled}{labelfont=normalfont,labelsep=colon,strut=off} 
\floatstyle{ruled}
\newfloat{listing}{tb}{lst}{}
\floatname{listing}{Listing}
\title{HALO: \underline{H}ardware-\underline{a}ware quantization with \\ \underline{lo}w critical-path-delay weights for LLM acceleration}
\author {
    Rohan Juneja\textsuperscript{\rm 1},
    Shivam Aggarwal\textsuperscript{\rm 1},
    Safeen Huda\textsuperscript{\rm 2},
    Tulika Mitra\textsuperscript{\rm 1},
    Li-Shiuan Peh\textsuperscript{\rm 1}
}
\affiliations {
    \textsuperscript{\rm 1}National University of Singapore,
    \textsuperscript{\rm 2}OpenAI\\
    \{rohan, shivam, tulika, peh\}@comp.nus.edu.sg, saf@openai.com
}

\usepackage{bibentry}

\begin{document}

\maketitle
\begin{abstract}
Quantization is critical for efficiently deploying large language models (LLMs). Yet conventional methods remain hardware-agnostic, limited to bit-width constraints, and do not account for intrinsic circuit characteristics such as the timing behaviors and energy profiles of Multiply-Accumulate (MAC) units. This disconnect from circuit-level behavior limits the ability to exploit available timing margins and energy-saving opportunities, reducing the overall efficiency of deployment on modern accelerators. To address these limitations, we propose \textit{HALO}, a versatile framework for Hardware-Aware Post-Training Quantization (PTQ). Unlike traditional methods, \textit{HALO} explicitly incorporates detailed hardware characteristics, including critical-path timing and power consumption, into its quantization approach. \textit{HALO} strategically selects weights with low critical-path-delays enabling higher operational frequencies and dynamic frequency scaling without disrupting the architecture's dataflow. Remarkably, \textit{HALO} achieves these improvements with only a few dynamic voltage and frequency scaling (DVFS) adjustments, ensuring simplicity and practicality in deployment. Additionally, by reducing switching activity within the MAC units, \textit{HALO} effectively lowers energy consumption. Evaluations on accelerators such as Tensor Processing Units (TPUs) and Graphics Processing Units (GPUs) demonstrate that \textit{HALO} significantly enhances inference efficiency, achieving average performance improvements of 270\% and energy savings of 51\% over baseline quantization methods, all with minimal impact on accuracy.
\end{abstract}

\begin{links}
    \link{Code}{https://github.com/ecolab-nus/HALO}
\end{links}

\section{1.\hspace{0.5cm}Introduction}
Transformer-based large language models (LLMs) have grown exponentially, increasing 100-fold every two years, far outpacing the 3.1× improvements in hardware. This widening gap has made inference increasingly costly, as seen with models like LLaMA (65 billion parameters) and GPT-4 (1.76 trillion parameters), which require vast computational resources. Quantization, which reduces model size and cost by lowering bit-width, is crucial for efficiency. However, implementing quantization is challenging because of the diverse, fragmented landscape of hardware accelerators, each optimized for specific quantization techniques and data types. Existing quantization techniques fail to account for these diverse hardware, highlighting the need for adaptable strategies that can optimize model efficiency across different accelerator architectures by leveraging circuit-level characteristics.
\begin{figure}[t!]
	\scriptsize
	\centering
	\includegraphics[width=1.05\columnwidth]{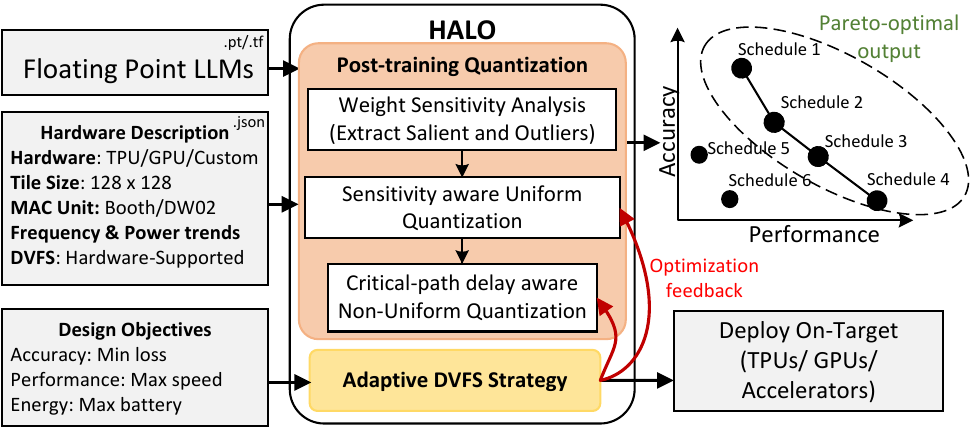}
	\caption{HALO quantization framework, using architectural details to yield Pareto-optimal trade-offs for diverse deployments.} 
	\label{fig:framework}
\end{figure}

Most existing quantization techniques~\cite{10705489} overly focus on reducing bit-width without considering hardware-specific factors, often treating critical components like Multiply-Accumulate (MAC) units as a black box.
In ML accelerators like Google's TPUs, MAC units used for matrix multiplications central to LLM inference, occupy 77-80\% of the chip area and account for 50-89\% of total power consumption~\cite{flextpu}. \textbf{Our key observation is that MAC units exhibit significant variability in performance and power characteristics, depending on specific weight patterns after quantization.}

Critical-path-delay in a hardware circuit refers to the time taken by the longest chain of logic operations, which determines the maximum speed at which the overall system can operate. In digital circuits, the path taken by a computation depends on the weight values being processed.
Through detailed circuit analysis, we discover that certain weight configurations enable shorter critical-paths, allowing the MAC units to operate at higher frequencies, whereas others result in longer critical-paths that constrain the system's overall operating speed, as illustrated in Fig.\ref{fig:frequency_trend} and Fig.\ref{fig:power_trend}. This variability highlights an unexplored opportunity to enhance inference performance and energy efficiency through more sophisticated, hardware-aware weight quantization strategies. Yet, current techniques remain oblivious to this crucial quantization-hardware interplay.

Hardware accelerators for LLMs, such as GPUs, use ~\emph{Dynamic Voltage and Frequency Scaling (DVFS)} to balance performance and power. NVIDIA GA100 GPUs, for instance, support up to 181 DVFS configurations~\cite{gpu_dvfs_181}. Quantization directly influences the critical-path-delays in MAC units, which in turn determines how well DVFS can be utilized to optimize both performance and energy efficiency. Our key insight is that by carefully selecting quantization levels that correspond to favorable critical-path-delays, we can enable higher operating frequencies while maintaining model accuracy.

We present \emph{HALO}, Hardware-Aware LOw critical-path delay quantization framework, illustrated in Fig.~\ref{fig:framework}, which unifies post-training quantization and DVFS into a single hardware-driven optimization strategy. At a high level, HALO first groups weights by their critical‑path-delay and then assigns each group the best voltage‑frequency setting to speed up ``fast” tiles and save energy on ``slow” ones. HALO accepts detailed hardware descriptions, such as MAC unit frequency and power profiles, supported DVFS configurations, and tile sizes, along with user-defined design goals for accuracy, performance, and energy efficiency. It outputs a set of Pareto-optimal quantized models, each paired with a corresponding DVFS schedule tailored for efficient deployment on target accelerators such as GPUs and TPUs. 

Below, we outline our main contributions.
\begin{itemize}
    \item We perform comprehensive evaluation of MAC circuit timing and energy variations, identifying weights that are critical for preserving model accuracy through gradient-based sensitivity profiling.
    \item We introduce HALO, a hardware-aware quantization method utilizing timing, energy, and weight sensitivity to boost LLM efficiency.
    \item We propose efficient integration techniques for HALO into existing hardware architecture such as TPUs and GPUs, enhancing performance while preserving accuracy.
\end{itemize}
Our integrated approach delivers significant benefits for LLM inference, achieving on average 270\% performance gains and 51\% energy savings. These improvements are driven by HALO's ability to co-optimize quantization levels with DVFS operating points, effectively bridging the gap between model compression and hardware adaptation.

\section{2.\hspace{0.5cm}Related Works}
\label{sec:relatedworks}
\subsubsection{Quantization Methods.}
Post-training quantization (PTQ) methods such as GPTQ~\cite{gptq} and AWQ~\cite{awq} aim to reduce LLM weights to 3 or 4 bits while preserving key precision. Other approaches like outlier-aware quantization~\cite{owq} and SqueezeLLM~\cite{squeezellm} differentiate between sensitive and non-sensitive weights to guide quantization. SmoothQuant~\cite{smoothquant} shifts quantization complexity to weights, enabling both weights and activations to be quantized to 8 bits. However, these methods operate at the algorithmic level and do not account for underlying hardware. HALO is the first framework to integrate circuit-level insights, considering timing and DVFS constraints, into the quantization process to improve efficiency without compromising LLM accuracy.

\subsubsection{Quantization Accelerators.}
GOBO~\cite{gobo} and OlAccel~\cite{olaccel} use mixed precision to preserve accuracy, while BitFusion~\cite{bitfusion} adapts precision based on workload needs. 
GOBO depends on full-precision compute units, and OlAccel encodes outliers using coordinate lists, both requiring custom accelerators. HALO, in contrast, is an orthogonal software-level technique that enhances efficiency without demanding hardware changes, making it suitable for deployment on current AI infrastructure (such as GPUs and TPUs), as well as future accelerators.
\section{3.\hspace{0.5cm}Background}
\label{section:motivation}
\begin{figure}[t!]
	\scriptsize
	\centering
	\includegraphics[width=1.05\columnwidth]{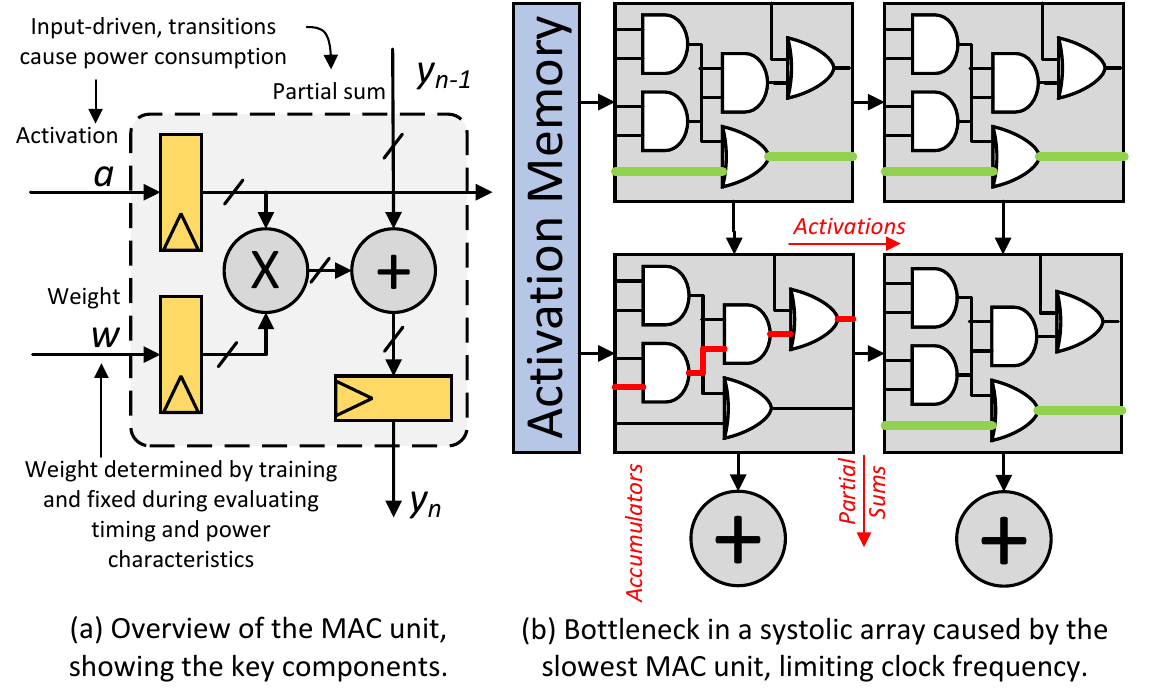}
	\caption{Impact of MAC unit on systolic array efficiency.} 
	\label{fig:motivation}
\end{figure}

The Multiply-Accumulate (MAC) unit is a fundamental component in AI accelerators, playing a significant role in both power consumption and area utilization. As illustrated in Fig.~\ref{fig:motivation}(a), each MAC unit features three input ports and two output ports. 
The unit operates by multiplying the weight $w$ with the activation $a$ to produce the product $wa$. This product is then added to the third input $y_{n-1}$, resulting in the updated partial sum $y_n$. 
\begin{figure}[t!]
	\centering
	\begin{tikzpicture}
    \begin{axis}[
        width=0.54\columnwidth, height=3.6cm,
        xlabel={Delay (in ps)},
        ylabel={Frequency},
        ymin=0, ymax=5500,
        xmin=150, xmax=280,
        xtick={175, 225, 275},
        ytick={0, 2500, 5000},
        title={\scriptsize{\shortstack{Quantized weight value 64,\\maximum delay: 265 ps}}},
        title style={yshift=-1ex},
        every axis plot/.style={thick, teal},
        clip=false
    ]
        \addplot[ybar, draw=none, fill=cadetblue, bar width=1.4pt] table[x=x, y=y, col sep=comma] {graphs/data/64_data.csv};

        \node[red, very thick, rotate=0] at (axis cs:265, 2250) {\Huge $\downarrow$};
    \end{axis}

    \hspace{-6cm}
    \begin{axis}[
        width=0.54\columnwidth, height=3.6cm,
        at={(10cm, 0)}, 
        xlabel={Delay (in ps)},
        ylabel={},
        ymin=0, ymax=700,
        xmin=370, xmax=540,
        xtick={400, 450, 500, 550},
        ytick={0, 250, 500},
        title={\scriptsize{\shortstack{Quantized weight value -127,\\maximum delay: 522 ps}}},
        title style={yshift=-1ex},
        every axis plot/.style={thick, teal},
        clip=false
    ]
        \addplot[ybar, draw=none, fill=cadetblue, bar width=1.1pt] table[x=x, y=y, col sep=comma] {graphs/data/127_data.csv};

        \node[red, very thick, rotate=0] at (axis cs:522, 300) {\Huge $\downarrow$};
    \end{axis}
\end{tikzpicture}
    \caption{Delay profiles for two weight values. Arrows indicate the maximum delay for each weight across all activations.} 
	\label{fig:delay_profile}
\end{figure}
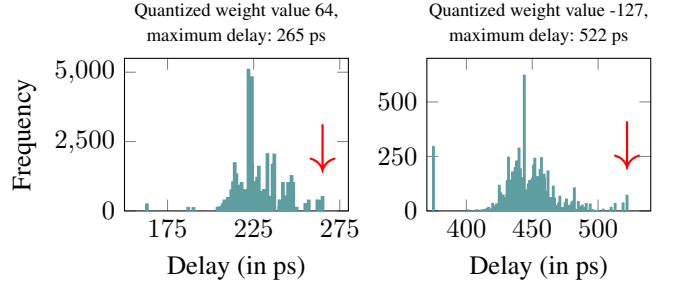

The timing characteristics of a MAC unit are heavily influenced by the specific weight values, as they affect the worst-case critical-path-delays, ultimately constraining the operating frequency. 
Using Synopsys tools~\cite{synopsys_primetime}, we perform static timing analysis on the 8-bit MAC unit commonly deployed in modern TPUs and GPUs.
Fig.~\ref{fig:delay_profile} illustrates the timing profile for two quantized weights, 64 and -127, with the x-axis representing delay and the y-axis showing the frequency of this delay across all activation transitions. The weight value 64 achieves an operating clock frequency of 3.7 GHz, while -127 is limited to 1.9 GHz. 
Certain bit patterns reduce the number of active signal paths, shortening critical-paths and resulting in faster processing for specific weight values.
\begin{figure}[t!]
	\centering
	\begin{tikzpicture}
\begin{groupplot}[
  group style={
    group name=my plots,
    group size=1 by 1,
    xlabels at=edge bottom,
    xticklabels at=edge bottom,
    vertical sep=7pt
  },
  xtick={-128, -112, -96, -80, -64, -48, -32, -16, 0, 16, 32, 48, 64, 80, 96, 112, 127}, 
  x tick label style={font=\scriptsize,rotate=45,anchor=east,text height=3pt},
  xlabel={Weight Values (8-bit Quantization)},
  ylabel={Achievable frequency\\(in GHz)},
  width=\columnwidth, height=3.2cm, 
  ymin=1.8
]

\nextgroupplot[
  width=\columnwidth, height=3.6cm,
  label style={align=center},
  bar width=0.80pt,
  ybar=0pt,
  ymin=1.8,
  enlarge x limits=0.04, 
]

\addplot[draw=none, fill=cadetblue]
    table [
    x=x,
    y=y
]{graphs/data/frequency_trend.data};

\end{groupplot}
\end{tikzpicture}
	\caption{Achievable frequency (GHz) for 8-bit quantized weight values from -128 to 127. Peaks indicate weights with lower critical-path-delays, allowing for higher operating frequencies.} 
	\label{fig:frequency_trend}
\end{figure}
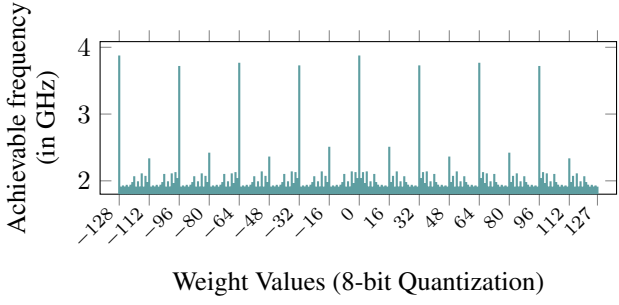

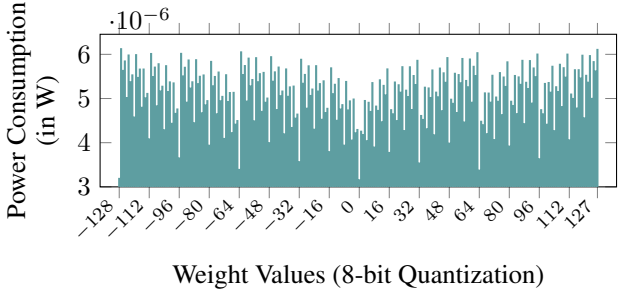
\begin{figure}[t!]
	\centering
	\begin{tikzpicture}
\begin{groupplot}[
  group style={
    group name=my plots,
    group size=1 by 1,
    xlabels at=edge bottom,
    xticklabels at=edge bottom,
    vertical sep=7pt
  },
  xtick={-128, -112, -96, -80, -64, -48, -32, -16, 0, 16, 32, 48, 64, 80, 96, 112, 127}, 
  x tick label style={font=\scriptsize,rotate=45,anchor=east,text height=3pt},
  xlabel={Weight Values (8-bit Quantization)},
  ylabel={Power Consumption\\(in W)},
  width=\columnwidth, height=5cm, 
  ymin=1.8
]

\nextgroupplot[
  width=\columnwidth, height=3.6cm,
  label style={align=center},
  bar width=0.80pt,
  ybar=0pt,
  ymin=0.000003,
  enlarge x limits=0.04, 
]

\addplot[draw=none, fill=cadetblue]
    table [
    x=x,
    y=y
]{graphs/data/power_trend.data};

\end{groupplot}
\end{tikzpicture}
	\caption{Power consumption (in Watts) for 8-bit quantized weight values ranging from -128 to 127, where lower values reflect decreased power usage due to reduced switching activity.} 
	\label{fig:power_trend}
\end{figure}

Fig.~\ref{fig:frequency_trend} shows the achievable operating frequency, based on the maximum delay for each weight value across activation transitions, while Fig.~\ref{fig:power_trend} depicts the corresponding power consumption for an 8-bit integer MAC unit. Active power consumption is determined by switching activity, which fluctuates with different weight values. Notably, weights associated with shorter critical-path-delays exhibit lower power consumption, offering potential energy savings. This observed correlation between timing and power characteristics reveals opportunities for optimizing both frequency and energy by strategically selecting weight values for model inference.

The behavior of individual MAC units is crucial in systolic arrays, where performance relies on synchronized operation driven by a global clock, ensuring seamless dataflow across all units. Each MAC unit operates in lockstep to ensure seamless data flow at every clock cycle. 
This global synchronization, necessary for maintaining data alignment, makes the slowest unit a performance bottleneck. 
Fig.~\ref{fig:motivation}(b) illustrates the MAC units arranged in a systolic array, emphasizing how the critical-path of the slowest unit restricts the clock frequency. To address this limitation, HALO quantizes weights while accounting for the frequency and energy trends of MAC units and subsequently detects the optimal DVFS configuration. 
{HALO restricts weights within each tile to the same or higher frequency class, enabling execution at the highest safe operating point without violating critical-path timing. All MAC units within a tile are clocked uniformly, ensuring deterministic behavior and preserving the synchronous dataflow of the architecture.}
This integrated approach enhances both performance and energy, providing substantial advancement over traditional methods.
\begin{figure*}[t!]
	\scriptsize
	\centering
	\includegraphics[width=\textwidth]{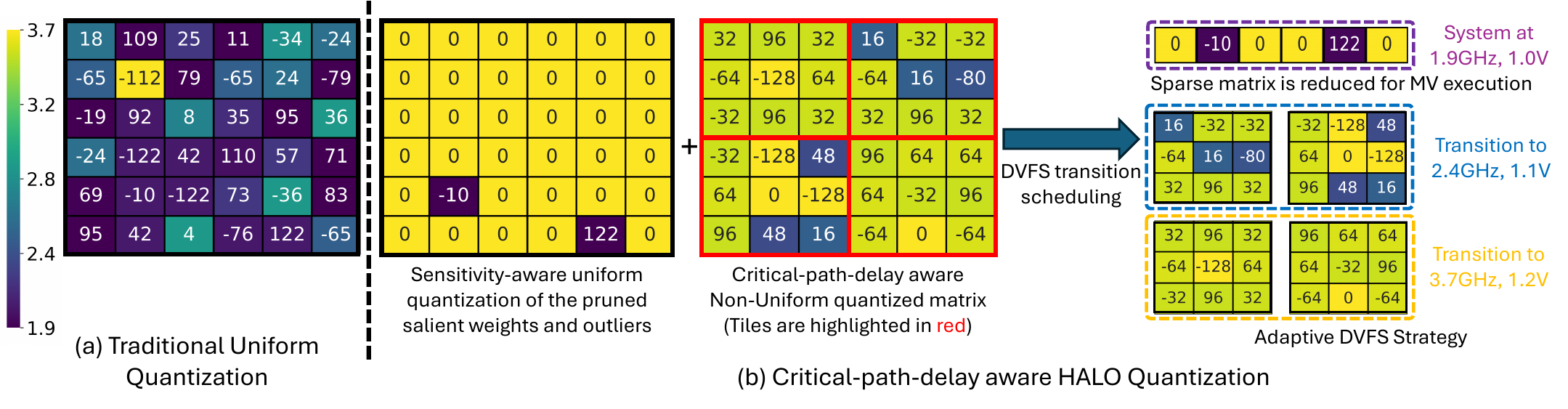}
	\caption{Illustrative example comparing traditional and HALO quantization for 3x3 tile size, highlighting HALO’s awareness of critical-path-delay. Only weight quantization is shown; activations are uniformly quantized to 8 bits in both cases. Tile distribution across frequency classes is identical here but depends on user-defined goals for accuracy, performance, and energy.} 
	\label{fig:halo_example}
\end{figure*}
\section{4.\hspace{0.5cm}Quantization Framework}
\label{section:quantization}
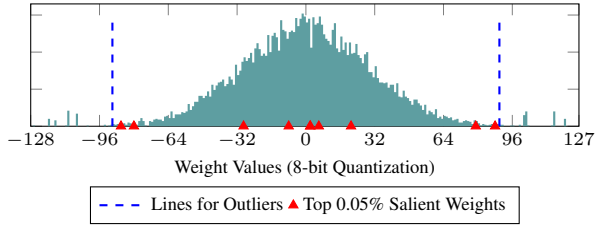
\begin{figure}[h!]
	\scriptsize
	\centering
	\begin{tikzpicture}
    \begin{axis}[
        width=1.05\columnwidth, height=3.2cm,
        title={},
        xlabel={},
        ylabel={},
        yticklabels={},
        xtick={-128, -96, -64, -32, 0, 32, 64, 96, 127},
        xticklabel style={/pgf/number format/fixed},
        xlabel={Weight Values (8-bit Quantization)},
        ymin=0,
        ymax=165,
        legend columns=2,
    legend style={fill=white, at={(0.5,-0.8,1.15)}, anchor=south, font=\scriptsize},
        enlargelimits=false,
    ]
        \addplot+[ybar interval, mark=none, draw=none, fill=cadetblue] 
            table[x=x, y=y, col sep=comma] {graphs/data/data.csv};
        
        \addplot[blue, thick, dashed] coordinates {(-90, 0) (-90, 150)};
        \addlegendentry{};
        \addplot[blue, thick, dashed] coordinates {(90, 0) (90, 150)};
        
        \pgfplotstableread[col sep=comma]{graphs/data/sensitive_values.csv}\sensitivevalues
        \addplot[only marks, mark=triangle*, mark size=2pt, color=red]
            table[x=x, y=y] {\sensitivevalues};
        
        \legend{,,Lines for Outliers,  Top 0.05\% Salient Weights}
    \end{axis}
\end{tikzpicture}
	\caption{Distribution highlighting sensitive weights important for accuracy.} 
	\label{fig:weight_distribution}
\end{figure}
Fig.~\ref{fig:halo_example} illustrates HALO’s quantization strategy using a simplified 3×3 tile. Traditional quantization (Fig.~\ref{fig:halo_example}a) applies uniform 8-bit precision to all weights, ignoring MAC-level timing and power variation, which restricts the use of voltage and frequency scaling. In contrast, HALO preserves high precision for salient and outlier weights, which are extracted and encoded as a sparse matrix. The remaining weights are quantized using a critical-path-delay aware scheme and organized into tiles based on their MAC timing characteristics (Fig.~\ref{fig:halo_example}b). This sparse matrix is reduced to matrix-vector form for efficient execution on dedicated SpMV units. {All MAC units within a tile operate at the same frequency, and tiles assigned to the same frequency class are executed together, requiring only a single DVFS transition per class.} This design preserves architectural dataflow while minimizing runtime control overhead.

\begin{algorithm}[t!]
\caption{Quantization Framework}
\label{quantization}
{\footnotesize
\begin{algorithmic}[1]
\Require calibration dataset $X$, pre-trained weight matrix $W$, gradient $G$
\Require tile size $t$, quantile threshold $k$, target frequencies $f_1$, $f_2$
\Ensure Quantized weight matrix $W_q$

\State $W_s, S \leftarrow \text{ExtractSalientValues}(W, G)$ \Comment{Isolate values with high saliency}
\State $W_o, O \leftarrow \text{ExtractOutliers}(W_s)$ \Comment{Separate outlier weights}
\State $W_{s,o}^{q} \leftarrow \text{Quantize}(W_{s} + W_{o})$  \Comment{Quantize outliers and salient weights}

\State $W_t \leftarrow 
\text{ReshapeIntoTiles}(\text{PadMatrix}(W_o, t), t)$ \Comment{Tile reshaping}

\State $\Lambda_{T_k} \leftarrow \text{CalculateTileSensitivities}(G)$ \Comment{Compute sensitivity for each tile}
\State $M_l, M_h \leftarrow \text{CreateMasks}(M, \text{ComputeAdaptiveK}(\Lambda_{T_k}, k))$ \Comment{Classify tiles as low or high sensitivity}
\State $W_{l,i}, W_{h,i} \leftarrow W_{t,i} \odot M_{l,i}, W_{t,i} \odot M_{h,i}$ \Comment{Apply masks}
\State $W_{l,i} \leftarrow \text{Quantize}(W_{l,i}, f_1)$ \Comment{Quantize low-sensitivity tiles}
\State $W_{h,i} \leftarrow \text{Quantize}(W_{h,i}, f_2)$  \Comment{Quantize high-sensitivity tiles}
\State $W_q \leftarrow W_{l,i}, W_{h,i}, W_{s,o}$
\State \Return $W_q$
\end{algorithmic}
}
\end{algorithm}

Building on this design, our quantization framework introduces a timing-aware method for LLM inference, detailed in Algorithm~\ref{quantization}. 
It prioritizes weights with low critical-path-delays to reduce latency while maintaining model fidelity. The adaptive method operates across quantization levels and layer sensitivities, optimizing performance by focusing on weights most critical to overall efficiency. The framework consists of three key stages: \circled{1} sensitivity-aware uniform quantization to identify and retain critical weights \textit{(Lines 1–3)}, \circled{2} critical-path-delay aware non-uniform quantization to optimize remaining weights for hardware efficiency \textit{(Lines 4–10)}, and \circled{3} adaptive DVFS, which assigns optimal frequency levels based on tile timing characteristics to maximize performance and energy efficiency.

\subsection{Stage 1: Sensitivity-aware Uniform Quantization}
The framework begins with a sensitivity analysis of model weights, identifying weights values that can tolerate quantization without significantly affecting accuracy, as outlined in Algorithm 1. 
The framework initially separates \textit{outliers} (outside the blue lines) and \textit{salient weights} (in red) from normal values, as shown in Fig.~\ref{fig:weight_distribution}.

\subsubsection{Outliers \& Salient Weights}: We incorporate outlier removal to manage extreme weight values based on inter-quartile range scaling. To compute outliers in the weight distribution, we employ the 3$\sigma$ rule~\cite{olive}. Outliers are identified as values lying beyond three standard deviations from the mean. 


From the normal values obtained after this distribution, we rely on Taylor series expansion to estimate the most salient weights in the model. Following~\cite{squeezellm}, we use an approximation to the Hessian \(H \) based on the Fisher information matrix \( F \), which can be calculated over a sample dataset \( D \) as
\begin{equation}
F = \frac{1}{|D|} \sum_{d \in D} g_d g_d^{\top},
\label{eq:fisher}
\end{equation}
where \( g \) is the gradient and \( H \approx F \). This only requires computing the gradient for a set of samples. For each weight tensor \( W \), the weight sensitivity is computed as \(\Lambda_{W} = F \). Weights with higher \( \Lambda_{W} \) values are considered more salient due to their significant impact on the model's output. We preserve the top 0.05\% of the weights based on this criterion. Cumulatively, both outliers and extremely salient weight values correspond to less than 0.5\% of the total weight values. For this reason, we handle these weight values separately and apply per-channel quantization for this set of weight values, isolating them to maintain model precision.

\subsection{Stage 2: Critical-path-delay aware Non-Uniform Quantization} 
\label{section:crit_quant}
We leverage non-uniform quantization to more efficiently map the distribution of weights to specific values that reduce the critical-path-delays, thereby optimizing frequency and energy consumption.

\subsubsection{Tile-Based Sensitivity Analysis:} To optimize the model for efficient inference on hardware, the weight tensors are divided into fixed-size tiles (\(128 \times 128\) by default). Specifically, the sensitivity of each tile is evaluated 
as the sum of the absolute values of the gradients for each tile, normalized by the size of the tile, based on Eq.~\ref{eq:fisher}. For a given $k$th tile \( T_k \), we compute a \textit{per-tile sensitivity score} \( \Lambda_{T_k} \) using a diagonal approximation of the Fisher information matrix:
\begin{equation}
\Lambda_{T_k} = \frac{\sum_{i,j} g_{k,i,j}^2}{\text{tile\_rows} \times \text{tile\_cols}}
\end{equation}

where \( g_{k,i,j} \) denotes the gradient of the loss with respect to each weight in the \( k \)-th tile, and \( \text{tile\_rows} \times \text{tile\_cols} \) represents the total number of elements within the tile. This score captures the average Fisher information across all weights in the tile, providing a quantitative measure of the tile's sensitivity in relation to its influence on the model's output.

\subsubsection{Tile Sensitivity Mapping}: 
To balance hardware efficiency and model accuracy, tiles in each layer are classified as \underline{low-sensitive} or \underline{high-sensitive} based on their relative importance. Determining a fixed sensitivity threshold for each layer is challenging, as weight distributions vary significantly across layers. To address this, we employ a dynamic tile sensitivity mapping strategy that adapts to the cumulative sensitivity distribution of each layer.

The process starts by computing the sensitivity of all tiles in a given layer, derived as the normalized sum of absolute gradient magnitudes within each tile. Sensitivities are then sorted in descending order to rank tiles by importance. A cumulative sum of these sorted sensitivities is then normalized against the total layer sensitivity, generating a cumulative distribution curve from 0 to 1.

The mapping threshold \(k \) is derived from this curve and represents the fraction of tiles classified as low-sensitive, ensuring a specified percentage of total sensitivity (e.g., 95\%) is retained. Tiles contributing most to overall sensitivity are marked high-sensitive, while the rest are classified as low-sensitive. Mathematically, \(k \) is the ratio of the index where cumulative sensitivity exceeds the threshold to the total number of tiles, defaulting to 1.0 if no such index exists.

Once \(k \) is determined, boolean masks separate tiles into low- and high-sensitivity categories. Low-sensitivity tiles are quantized more aggressively, while high-sensitivity tiles retain higher precision to preserve performance. The adaptive quantization and computation flow based on the DVFS characteristics are described next.
\subsection{Stage 3: Adaptive DVFS Strategy}
\subsubsection{DVFS for Outliers and Salient features:}
\begin{itemize}
    \item {Packaging of Salient and Outlier Weights:}
Salient and outlier weights, exhibiting extreme sparsity, are packaged for efficient computation using a Sparse Matrix-Vector Multiplication (SpMV) engine. The hypersparse weight matrix is compactly stored with a value vector $val$ for non-zero elements and an index vector $idx$ for column positions, reducing memory usage and accelerating computations. The matrix-vector multiplication $A \times b$ for a dense vector $b \in \mathbb{R}^k$ is performed as:
\[\text{res}[i] = \text{val}[i] \times b[\text{idx}[i]], \quad \text{for } i = 0, 1, \ldots, m - 1,\]
where $res$ is the result vector, efficiently leveraging the matrix’s sparsity.
\item {DVFS Configuration Selection:}
These uniformly quantized hypersparse weights span the entire 8-bit range, necessitating DVFS settings that respect the critical-path-delay for all possible weight values. Guided by MAC characteristic trends, the optimal voltage-frequency \( (V, f) \) point is selected to minimize energy consumption while ensuring timing constraints are met:
\hspace{-0.2cm}\[(V, f) = \arg\min_{(V_i, f_i)} \left[ E(V_i, f_i) \right], \text{given} (1/f_i) \geq \text{Critical-Path}\]
This approach ensures efficient system performance without sacrificing model fidelity.
\end{itemize}

\subsubsection{DVFS for High- and Low-Sensitivity Weights:} For non-uniformly quantized weights on the systolic array, weight tensors are divided into $128\times128$ tiles, with DVFS settings determined by the tile sensitivity mapping strategy.
This tile size aligns with the architecture of modern systolic arrays in TPUs and is critical for balancing the tradeoff between accuracy and performance, optimizing energy efficiency while preserving precision.

\begin{table}[h!]
\centering
\resizebox{\columnwidth}{!}{
\begin{tabular}{c|c}
\hline
\textbf{Hardware}       & \textbf{DVFS Levels (Voltage, Frequency)} \\ \hline
GPU                     & (0.9 V, 1.5 GHz), (1.0 V, 2.0 GHz), (1.1 V, 2.8 GHz) \\
Systolic Array (TPU)    & (1.0 V, 1.9 GHz), (1.1 V, 2.4 GHz), (1.2 V, 3.7 GHz)\\ \hline
\end{tabular}
}
\caption{Assumed DVFS levels for GPUs and Systolic Arrays.}
\label{tab:dvfs_levels}
\end{table}

The assumed DVFS levels for both GPUs and systolic arrays are summarized in Table~\ref{tab:dvfs_levels}. We base our DVFS levels on practical hardware parameters: for GPUs, using NVIDIA's publicly available maximum clock frequency of 2.8 GHz~\cite{gpu_max_frequency}, and for systolic arrays (similar to TPUs), deriving levels from MAC characteristics.

\begin{itemize}
    \item {High-Sensitivity Tiles:} These tiles are composed of critical weights that are crucial for maintaining model accuracy. As described in the background section, the 
    MAC unit handles 16 high-sensitivity weights, operating at a frequency of 2.4 GHz. These tiles are exclusively made up of these 16 values to ensure precision. The execution is performed at specific DVFS settings ($V_i$,$f_i$), where $1/f_i \ge$ the critical delay associated with these weights. While this configuration may lead to higher energy consumption, it significantly boosts tile performance by overclocking the accelerator's global clock to meet the precise timing requirements of these weights.
    \item {Low-Sensitivity Tiles:} These tiles are subjected to aggressive overclocking to maximize performance. They contain only 9 weights, each capable of operating at frequencies up to 3.7 GHz. The DVFS settings are fine-tuned to respect the critical-path-delay only for these 9 weights. Despite the high overclocking, the energy increase remains incremental, as switching occurs primarily along the shortest paths in the circuit.
\end{itemize}

\subsubsection{DVFS Transition Scheduling and Overhead Minimization:}
HALO reduces DVFS overhead by grouping tiles that share the same frequency to run together. Each frequency level is activated once per group, limiting the number of voltage and clock transitions to just a few, each lasting tens of nanoseconds~\cite{dvfs_ns} to a few microseconds, significantly shorter than typical LLM inference times (Llama2-13B requires 53 ms, and OPT-30B exceeds 120 ms). Since most models use only two or three frequency levels, the total number of transitions remains low, even in large-scale deployments.

This grouping is a scheduling optimization. Quantization and frequency assignments are determined offline, and execution order does not affect dataflow, timing, or accuracy. Clocking remains uniform within each tile, ensuring that the functional behavior and numerical correctness of the model are preserved.

By leveraging MAC characteristics, this approach achieves an efficient trade-off between processing speed and energy consumption, optimizing the use of the accelerator’s resources.
\section{5.\hspace{0.5cm}Evaluation}
\label{section:evaluation}
This section evaluates the accuracy of LLM models with HALO quantization and demonstrates its speedup and energy efficiency compared to systolic arrays and GPUs.

\subsection{Implementation Details}
\subsubsection{Datasets and Models.}
We evaluate the effectiveness of HALO using various models, including LLaMA2~\cite{llama2} and OPT~\cite{opt} family of models.
We conduct language modeling evaluation using the C4~\cite{c4} and WikiText2~\cite{wikitext} datasets. We report \textit{perplexity} as the measure of the performance of the model. 

\subsubsection{Hyperparameters \& Baselines.} We evaluate our approach against state-of-the-art quantization methods, including Round-To-Nearest (RTN) quantization, SmoothQuant~\cite{smoothquant}, GPTQ~\cite{gptq}, and ZeroQuant~\cite{zeroquant, zeroquant_v2}, under varying weight precision while keeping activations fixed at 8 bits. Quantization is applied to computationally intensive operators such as attention and linear layers, with per-token static quantization used for activations. We evaluate HALO using tile sizes ranging from 128$\times$128 to 32$\times$32 to study tradeoffs between performance and model fidelity.  We provide more hyperparameter details in the \textit{Supplementary Material}. 

\subsubsection{Hardware Setup.} To evaluate the systolic array's performance, we develop a custom simulator and implement the design in SystemVerilog with support for global DVFS control. The design is synthesized using 22nm technology and verified through behavioral simulation. For GPU results, we extend AccelSim~\cite{accelsim} to model the NVIDIA 2080 Ti with DVFS settings from Table~\ref{tab:dvfs_levels}, and estimate energy using AccelWattch~\cite{accelwattch} and GPUWattch~\cite{gpuwattch}.

\subsection{Accuracy Results}
\begin{table*}[h!]
\centering
\resizebox{\textwidth}{!}{
\begin{tabular}{clcccc|cccc}
\hline
\multirow{2}{*}{PPL$\downarrow$}  && \multicolumn{4}{c|}{WikiText} & \multicolumn{4}{c}{C4} \\
 && Llama2-7B& Llama2-13B& OPT-1.3B& OPT-30B& Llama2-7B& Llama2-13B& OPT-1.3B& OPT-30B\\\hline
Ideal& FP16& 5.47& 4.95& 14.72& 9.56& 7.52& 6.98& 16.96& 11.84\\\hline
\multirow{3}{*}{\shortstack{RTN\\WxA8}}  &x = 8& 5.58& 4.94& 15.23& 9.89& 7.69& 7.04& 17.65 & 12.35\\
&x = 4& 7.36& 5.47& 81.23& 3717.12& 10.23& 7.71& 76.94& 4162.58\\
&x = 3& 19480.51& 2552.73& 13477.29& 9122.24& NaN& 1951.28& 6719.15& 6136.87\\\hline
\multirow{3}{*}{\shortstack{SmoothQuant~\cite{smoothquant}}}  &x = 8& 5.51& 4.93& 14.76& 9.61& 7.56& 7.04& 16.52& 12.03\\
&x = 4& 7.26& 5.64&18.47& 14.44& 10.16& 8.00& 20.35& 31.10\\
&x = 3& 8406.97& 572.53& 3780.82& 1074.64& NaN& 1145.67&  1188.41& 1116.33\\\hline
{\shortstack{GPTQ~\cite{gptq}* WxA8}}&x = 4& $-$ & $-$ & 15.75 & 12.79 & $-$ & $-$ & 15.93 & 24.14\\\hline
{\shortstack{ZQ-Local~\cite{zeroquant}* WxA8}}&x = 4& $-$ & $-$ & 15.98  & 11.69& $-$ & $-$ & 16.20 & 18.96\\\hline
{\shortstack{ZQ-Global~\cite{zeroquant}* WxA8}}&x = 4& $-$ & $-$ & 15.77 & 11.80& $-$ & $-$ & 15.83 & 13.41\\\hline
\multirow{3}{*}{\shortstack{HALO (Tile=128)}}  &Perf-opt (BW)& 6.37 (3.03)& 5.47 (3.03)& 16.92 (3.08)& 9.95 (3.04)& 8.87& 7.78& 18.43& 12.24\\
&Acc-opt (BW)& 5.94 (3.88)& 5.20 (3.80)& 15.59 (3.96)& 9.71 (3.75)& 8.23& 7.40& 17.29& 12.05\\
&Bal (BW)& 6.01 (3.75)& 5.44 (3.06)& 16.06 (3.82)& 9.84 (3.42)& 8.34& 7.73& 17.68& 12.13\\
HALO (Tile=64)&Bal (BW)& 5.89 (3.62)& 5.31 (3.05)& 15.82 (3.64)& 9.76 (3.40)& 8.31& 7.50& 17.44& 12.10\\
HALO (Tile=32)& Bal (BW)& 5.63 (3.33)& 5.01 (3.03)&  15.56 (3.50)& 9.62 (3.24)& 8.04& 7.19& 17.08&12.05\\\hline
\end{tabular}
}
\caption{LLM accuracy on WikiText~\cite{wikitext} and C4~\cite{c4}, measured by perplexity with a sequence length of 2048. Lower values indicate higher accuracy. For HALO, approximate weight bit-width values are reported in brackets alongside perplexity.\\
\textsuperscript{*The results for these techniques are taken from this related work~\cite{zeroquant_v2}, and are reported only for models where results are available.}}
\label{tab:llm}
\end{table*}

We evaluate HALO against FP16 and WxA8 integer quantization schemes, where activations are quantized to 8 bits and weights to 8, 4, or 3 bits. We compare HALO to RTN, which loses accuracy at lower precisions, and to SmoothQuant, which shifts sensitivity from activations to weights. As shown in Table~\ref{tab:llm}, HALO closely matches FP16 accuracy, with perplexity degradation under 0.5 for most models using the \textit{acc-opt} and \textit{bal} variants. While RTN-W8A8 and SmoothQuant-W8A8 slightly outperform HALO in a few cases, their performance drops sharply at lower bit widths. We include W3A8 and W4A8 baselines as they align with the quantization levels used by HALO. MAC unit delay characteristics allow HALO to group weights into sets of 9 and 16 values, making these baselines relevant for comparison. \textbf{Advanced schemes like GPTQ, ZQ-Local, and ZQ-Global perform well on smaller models but fail to match the accuracy of HALO’s \textit{acc-opt} and show increasing perplexity with model size, where HALO remains stable.}

To compare heterogeneous quantization schemes, we compute effective bit-width as the weighted average across layers: $B_{\rm eff} = \sum_i P_i b_i$, with $P_i$ denoting the fraction of parameters at $b_i$ bits. Across different tile sizes, HALO maintains accuracy under aggressive quantization. \textbf{For both LLaMA2 and OPT models, HALO preserves performance better as models scale, highlighting the advantages of sensitivity-aware quantization and fine-grained tile selection.} This consistency underscores HALO’s suitability for large-scale deployments without compromising numerical robustness.

\subsection{Systolic Array Performance and Energy}
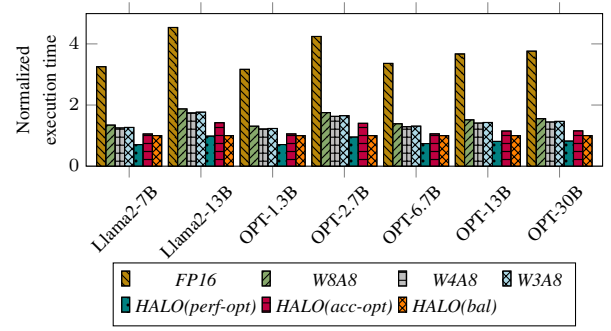
\begin{figure}[h!]
	\scriptsize
	\centering
	\begin{tikzpicture}
\begin{groupplot}[
  group style={
    group name=my plots,
    group size=1 by 1,
    xlabels at=edge bottom,
    xticklabels at=edge bottom,
    vertical sep=7pt
  },
  xtick=data,
  x tick label style={font=\scriptsize,rotate=45,text height=3pt]},
  xticklabels from table={graphs/data/performance_sys.data}{Model},
  legend style={fill=white, at={(0.5,1.05,1.15)}, anchor=south, font=\scriptsize},
]

\nextgroupplot[
  width=\columnwidth, height=3.6cm,
  label style={align=center},
  ylabel={Normalized\\execution time},
  bar width=3.5pt,
  ybar=0pt,
  ymin=0,
  typeset ticklabels with strut,
  legend columns=4,
  legend style={at={(0.5,-1.05,1.15)}, anchor=south},
]

\addplot[fill=darkgoldenrod, postaction={pattern=north west lines}]
table[
  x expr = \coordindex,
  y = FP16,
]{graphs/data/performance_sys.data};
\addlegendentry{\em FP16}

\addplot[fill=asparagus, postaction={pattern=north east lines}]
table[
  x expr = \coordindex,
  y = INT8(W8/A8),
]{graphs/data/performance_sys.data};
\addlegendentry{\em W8A8}

\addplot[fill=lightgray, postaction={pattern=grid}]
table[
  x expr = \coordindex,
  y = INT4(W4/A8),
]{graphs/data/performance_sys.data};
\addlegendentry{\em W4A8}

\addplot[fill=lightblue, postaction={pattern=crosshatch}]
table[
  x expr = \coordindex,
  y = INT3(W3/A8),
]{graphs/data/performance_sys.data};
\addlegendentry{\em W3A8}

\addplot[fill=teal, postaction={pattern=dots}]
table[
  x expr = \coordindex,
  y = SystolicX_perf,
]{graphs/data/performance_sys.data};
\addlegendentry{\em HALO(perf-opt)}

\addplot[fill=purple, postaction={pattern=horizontal lines}]
table[
  x expr = \coordindex,
  y = SystolicX_acc,
]{graphs/data/performance_sys.data};
\addlegendentry{\em HALO(acc-opt)}

\addplot[fill=orange, postaction={pattern=crosshatch}]
table[
  x expr = \coordindex,
  y = SystolicX_bal,
]{graphs/data/performance_sys.data};
\addlegendentry{\em HALO(bal)}

\end{groupplot}

\end{tikzpicture}
	\caption{Normalized execution time across quantization methods. Lower values denote faster execution, emphasizing HALO's efficiency.} 
	\label{fig:performance_sys}
\end{figure}

\begin{figure}[h!]
	\scriptsize
	\centering
	\pgfplotsset{compat=1.17}
\usepgfplotslibrary{groupplots} 

\begin{tikzpicture}
\begin{groupplot}[
    group style={
        group name=my plots,
        group size=2 by 2,
        xlabels at=edge bottom,
        ylabels at=edge left,
        horizontal sep=0.8cm,
        vertical sep=0.7cm
    },
    width=5cm, height=3cm,
    legend pos=north west,
    legend cell align={left},
    legend style={font=\small, at={(-0.8,-0.6,-0.2)},anchor=north west,legend columns=3},
    grid=both,
    xtick={1.0, 1.1, 1.2, 1.3, 1.4}, 
    x label style={font=\small},
    y label style={font=\small}
]

\nextgroupplot[
    ylabel={\scriptsize{Perplexity}},
    xmin=1.0, xmax=1.5,
    ymin=5.94, ymax=6.4,
    title={Plot 1: Llama2-7b},
    title style={yshift=-2ex} 
]

\addplot[
    color=indianred,
    mark=*,
    mark options={fill=indianred},
    line width=1pt
] table[x=llama_7_x, y=llama_7_y] {graphs/data/pareto_frontier.data};

\addplot[
    only marks,
    mark=*,
    mark size=3pt,
    color=darkolivegreen,
    fill=darkolivegreen
] coordinates {
    (1.053911243,6.013976097)
};


\nextgroupplot[
    xmin=1.0, xmax=1.45,
    ymin=5.2, ymax=5.5,
    title={Plot 2: Llama2-13b},
    title style={yshift=-2ex} 
]

\addplot[
    color=indianred,
    mark=*,
    mark options={fill=indianred},
    line width=1pt
] table[x=llama_13_x, y=llama_13_y] {graphs/data/pareto_frontier.data};

\addplot[
    only marks,
    mark=*,
    mark size=3pt,
    color=darkolivegreen,
    fill=darkolivegreen
] coordinates {
    (1.4219069335851982, 5.445988655)
};

\nextgroupplot[
    xlabel={\scriptsize{Normalized Performance}},
    ylabel={\scriptsize{Perplexity}},
    xmin=1.0, xmax=1.43,
    ymin=11.1, ymax=11.6,
    title={Plot 3: OPT-1.3b},
    title style={yshift=-2ex} 
]

\addplot[
    color=indianred,
    mark=*,
    mark options={fill=indianred},
    line width=1pt
] table[x=opt_6_7_x, y=opt_6_7_y] {graphs/data/pareto_frontier.data};

\addplot[
    only marks,
    mark=*,
    mark size=3pt,
    color=darkolivegreen,
    fill=darkolivegreen
] coordinates {
    (1.169621678, 11.294034)
};

\nextgroupplot[
    xlabel={\scriptsize{Normalized Performance}},
    xmin=1.0, xmax=1.42,
    ymin=10.55, ymax=10.85,
    title={Plot 4: OPT-2.7b},
    title style={yshift=-2ex} 
]

\addplot[
    color=indianred,
    mark=*,
    mark options={fill=indianred},
    line width=1pt
]  table[x=opt_13_x, y=opt_13_y] {graphs/data/pareto_frontier.data};
\addlegendentry{Pareto-front}

\addplot[
    only marks,
    mark=*,
    mark size=3pt,
    color=darkolivegreen,
    fill=darkolivegreen
] coordinates {
    (1.108324123, 10.65611744)
};
\addlegendentry{Knee point}

\end{groupplot}
\end{tikzpicture}
	\caption{Normalized performance vs. perplexity, showing the knee point for optimal efficiency-accuracy trade-off.} 
	\label{fig:pareto_frontier}
\end{figure}
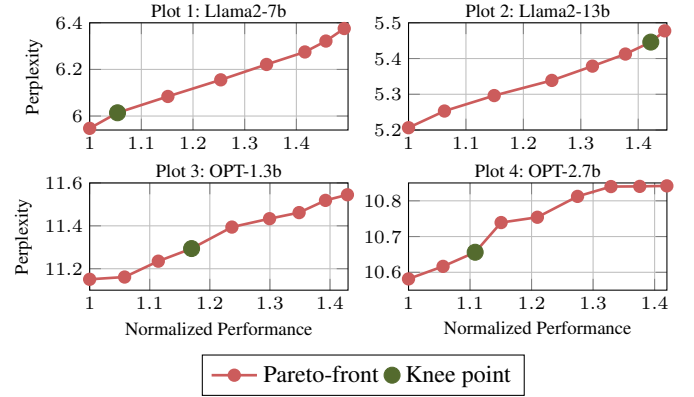
Fig.~\ref{fig:performance_sys} shows that HALO significantly reduces latency across all models, outperforming FP16 and uniform quantization schemes like W8A8, W4A8, and W3A8, which struggle with quantization-sensitive weights. HALO achieves 353\% speedup over FP16 and up to 87\% over other quantized baselines, while preserving higher accuracy than W4A8 and W3A8. These gains are consistent across different model sizes, highlighting HALO’s robustness under aggressive quantization. Execution time for handling outliers and salient weights remains below 1\% of total inference time, confirming that tile-level strategies drive most of the improvement. Fig.~\ref{fig:pareto_frontier} illustrates the tradeoff between performance and perplexity, with the \textit{bal} configuration corresponding to the knee point, representing an optimal balance for LLM workloads.

\begin{figure}[h!]
	\scriptsize
	\centering
	\input{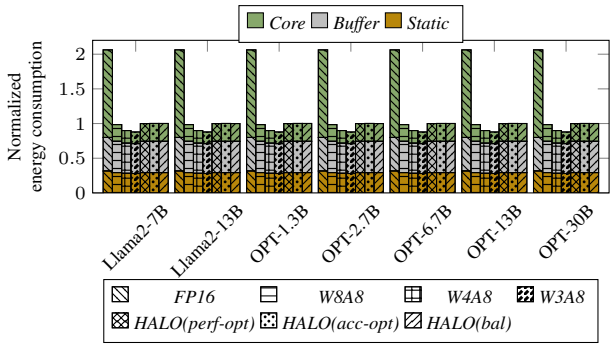}
	\caption{Normalized energy consumed across quantization methods.} 
	\label{fig:energy_sys}
\end{figure}
Fig.~\ref{fig:energy_sys} shows energy consumption split into static and dynamic components across core, buffer, and memory. FP16 is the most energy-intensive due to high-precision operations, while W8A8 saves some energy but lacks adaptability to model sensitivity. W4A8 scales better but shows an increase in energy with model size; W3A8 consumes the least energy, although with reduced accuracy. HALO improves energy efficiency by applying voltage and frequency scaling at tile granularity, guided by quantization sensitivity. Its energy use stays within 12\% of W3A8 and 10\% of W4A8, while offering much faster execution. 
\textbf{Even when operating at a higher voltage-frequency (VF) point, HALO does not exhibit any rise in total energy compared to W8A8.} This is because the increased dynamic power from higher frequency is offset by reduced switching activity within the MAC units, resulting in balanced energy consumption.
In systolic arrays, HALO clusters tiles with similar MAC delays, allowing synchronized frequency selection without slower tiles limiting throughput. \textbf{Since all tiles in a frequency group run together, the DVFS overhead is negligible and amortized across the group.} This lets HALO approach peak performance while preserving timing safety and energy proportionality.

\subsection{Impact of Tile Size on System Efficiency}
\label{sec:tile_sizes}
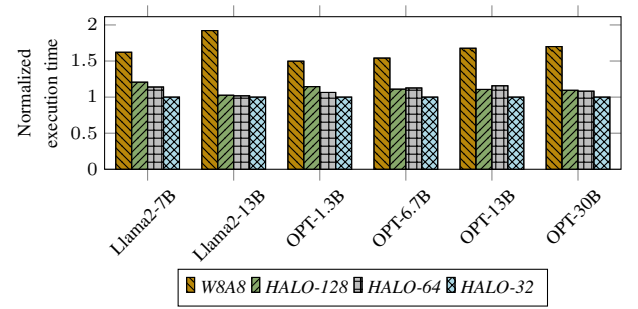
\begin{figure}[h!]
	\scriptsize
	\centering
	\begin{tikzpicture}
\begin{groupplot}[
  group style={
    group name=my plots,
    group size=1 by 1,
    xlabels at=edge bottom,
    xticklabels at=edge bottom,
    vertical sep=7pt
  },
  xtick=data,
  x tick label style={font=\scriptsize,rotate=45,text height=3pt]},
  xticklabels from table={graphs/data/performance_tile_sizes.data}{Model},
  legend style={fill=white, at={(0.5,1.05,1.15)}, anchor=south, font=\scriptsize},
]

\nextgroupplot[
  width=\columnwidth, height=3.6cm,
  label style={align=center},
  ylabel={Normalized\\execution time},
  bar width=6pt,
  ybar=0pt,
  ymin=0,
  typeset ticklabels with strut,
  legend columns=5,
  legend style={at={(0.5,-0.9,1.15)}, anchor=south},
]

\addplot[fill=darkgoldenrod, postaction={pattern=north west lines}]
table[
  x expr = \coordindex,
  y = int8,
]{graphs/data/performance_tile_sizes.data};
\addlegendentry{\em W8A8}

\addplot[fill=asparagus, postaction={pattern=north east lines}]
table[
  x expr = \coordindex,
  y = halo-128,
]{graphs/data/performance_tile_sizes.data};
\addlegendentry{\em HALO-128}

\addplot[fill=lightgray, postaction={pattern=grid}]
table[
  x expr = \coordindex,
  y = halo-64,
]{graphs/data/performance_tile_sizes.data};
\addlegendentry{\em HALO-64}

\addplot[fill=lightblue, postaction={pattern=crosshatch}]
table[
  x expr = \coordindex,
  y = halo-32,
]{graphs/data/performance_tile_sizes.data};
\addlegendentry{\em HALO-32}
\end{groupplot}

\end{tikzpicture}
	\caption{Normalized execution time of the systolic array for different tile sizes. HALO results are shown for the \textit{bal} configuration.} 
	\label{fig:tile_sizes}
\end{figure}
Fig.~\ref{fig:tile_sizes} shows HALO’s performance across varying tile sizes: HALO-128, HALO-64, and HALO-32, corresponding to $128\times128$, $64\times64$, and $32\times32$ tile dimensions. Smaller tiles, especially $32\times32$, improve performance by up to 15\% over $128\times128$ and 7\% over $64\times64$. This improvement stems from HALO’s ability to quantize a greater number of smaller tiles into higher frequency classes, thereby exploiting finer-grained control over timing and energy characteristics. Smaller tiles also enable the quantization engine to localize aggressive optimization decisions without being constrained by worst-case critical-paths across larger regions.

Table~\ref{tab:llm} shows that smaller tiles also lead to better perplexity in large models, confirming that finer granularity improves both hardware efficiency and model accuracy.\textbf{ By reducing synchronization overhead, smaller tiles let each MAC unit run closer to its ideal timing. These results highlight the importance of architecture-aware tile sizing for fully leveraging HALO’s quantization and DVFS strategy.}
These results emphasize the need for architecture-aware tile sizing strategies to fully exploit the benefits of HALO’s quantization and DVFS integration.

\subsection{GPU Performance and Energy}
\begin{figure}[h!]
	\scriptsize
	\centering
	\begin{tikzpicture}
\begin{groupplot}[
  group style={
    group name=my plots,
    group size=1 by 1,
    xlabels at=edge bottom,
    xticklabels at=edge bottom,
    vertical sep=7pt
  },
  xtick=data,
  x tick label style={font=\scriptsize,rotate=45,text height=3pt]},
  xticklabels from table={graphs/data/performance_gpu.data}{Model},
  legend style={fill=white, at={(0.5,1.05,1.15)}, anchor=south, font=\scriptsize},
]

\nextgroupplot[
  width=\columnwidth, height=3.6cm,
  label style={align=center},
  ylabel={Normalized\\execution time},
  bar width=5pt,
  ybar=0pt,
  ymin=0,
  typeset ticklabels with strut,
  legend columns=5,
  legend style={at={(0.5,-0.9,1.15)}, anchor=south, font=\scriptsize},
]

\addplot[fill=darkgoldenrod, postaction={pattern=north west lines}]
table[
  x expr = \coordindex,
  y = int8,
]{graphs/data/performance_gpu.data};
\addlegendentry{\em W8A8}

\addplot[fill=asparagus, postaction={pattern=north east lines}]
table[
  x expr = \coordindex,
  y = perf,
]{graphs/data/performance_gpu.data};
\addlegendentry{\em HALO (perf-opt)}

\addplot[fill=lightgray, postaction={pattern=grid}]
table[
  x expr = \coordindex,
  y = acc,
]{graphs/data/performance_gpu.data};
\addlegendentry{\em HALO (acc-opt)}

\addplot[fill=lightblue, postaction={pattern=crosshatch}]
table[
  x expr = \coordindex,
  y = bal,
]{graphs/data/performance_gpu.data};
\addlegendentry{\em HALO (bal)}
\end{groupplot}

\end{tikzpicture}
	\caption{Normalized execution time on GPUs.} 
	\label{fig:performance_gpu}
\end{figure}
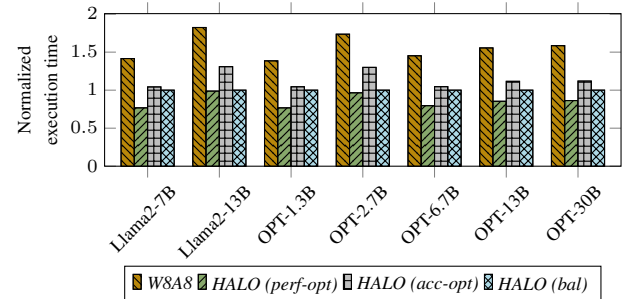
As shown in Fig.~\ref{fig:performance_gpu}, HALO consistently outperforms the \textit{W8A8} baseline on GPU by effectively managing weight sensitivity and aligning quantization levels with hardware capabilities. By selectively assigning frequency classes and applying dynamic voltage and frequency scaling, HALO reduces execution time across all evaluated models. Gains are especially notable in larger models, where quantization-induced bottlenecks are more severe. \textbf{As model sizes continue to grow, memory bandwidth and register pressure increase, making hardware-aware quantization and localized DVFS critical for maintaining throughput.}

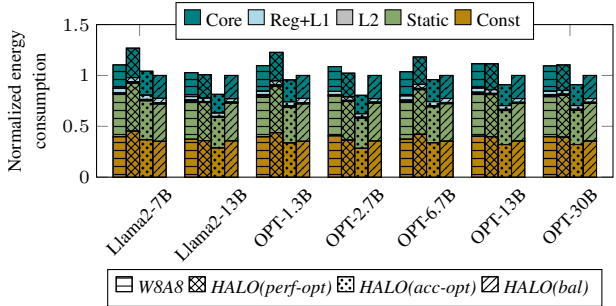
\begin{figure}[h!]
	\scriptsize
	\centering
	\begin{tikzpicture}
\begin{axis}[bar shift=-7.5pt,
    width=\columnwidth,, height=3.6cm,
    bar width=5pt,
    ybar stacked,
    ymin=0,
    ymax=1.5,
    ylabel={\shortstack{Normalized energy\\ consumption}},
    xtick=data,
    legend columns=4,
    legend columns=8,
    legend style={fill=white, at={(0.5,1.05,1.15)}, anchor=south, font=\scriptsize},
    reverse legend=true,
    xticklabels from table={graphs/data/energy_gpu.data}{Model},
    xticklabel style={rotate=45}
]

\addplot [fill=darkgoldenrod, postaction={pattern=horizontal lines, pattern color=black, line width=0.4pt}]
table[
    x expr = \coordindex,
    y = int8_const,
    ]{graphs/data/energy_gpu.data};

\addplot [fill=asparagus, postaction={pattern=horizontal lines, pattern color=black, line width=0.4pt}]
table[
    x expr = \coordindex,
    y = int8_static,
    ]{graphs/data/energy_gpu.data};

\addplot [fill=lightgray, postaction={pattern=horizontal lines, pattern color=black, line width=0.4pt}]
table[
    x expr = \coordindex,
    y = int8_l2_dram,
    ]{graphs/data/energy_gpu.data};

\addplot [fill=lightblue, postaction={pattern=horizontal lines, pattern color=black, line width=0.4pt}]
table[
    x expr = \coordindex,
    y = int8_reg_l1,
    ]{graphs/data/energy_gpu.data};

\addplot [fill=teal, postaction={pattern=horizontal lines, pattern color=black, line width=0.4pt}]
table[
    x expr = \coordindex,
    y = int8_core,
    ]{graphs/data/energy_gpu.data};
    
\end{axis}

\begin{axis}[hide axis, bar shift=-2.5pt,
    width=\columnwidth,, height=3.6cm,
    bar width=5pt,
    ybar stacked,
    ymin=0,
    ymax=1.5,
    xtick=data,
    legend columns=4,
    legend columns=8,
    legend style={fill=white, at={(0.5,1.05,1.15)}, anchor=south, font=\scriptsize},
    reverse legend=true,
    xticklabels from table={graphs/data/energy_gpu.data}{Model},
    xticklabel style={rotate=45}
]

\addplot [fill=darkgoldenrod, postaction={pattern=crosshatch, pattern color=black, line width=0.4pt}]
table[
    x expr = \coordindex,
    y = perf_const,
    ]{graphs/data/energy_gpu.data};

\addplot [fill=asparagus, postaction={pattern=crosshatch, pattern color=black, line width=0.4pt}]
table[
    x expr = \coordindex,
    y = perf_static,
    ]{graphs/data/energy_gpu.data};

\addplot [fill=lightgray, postaction={pattern=crosshatch, pattern color=black, line width=0.4pt}]
table[
    x expr = \coordindex,
    y = perf_l2_dram,
    ]{graphs/data/energy_gpu.data};

\addplot [fill=lightblue, postaction={pattern=crosshatch, pattern color=black, line width=0.4pt}]
table[
    x expr = \coordindex,
    y = perf_reg_l1,
    ]{graphs/data/energy_gpu.data};

\addplot [fill=teal, postaction={pattern=crosshatch, pattern color=black, line width=0.4pt}]
table[
    x expr = \coordindex,
    y = perf_core,
    ]{graphs/data/energy_gpu.data};
    
\end{axis}

\begin{axis}[hide axis, bar shift=2.5pt,
    width=\columnwidth,, height=3.6cm,
    bar width=5pt,
    ybar stacked,
    ymin=0,
    ymax=1.5,
    xtick=data,
    legend columns=4,
    legend columns=8,
    legend style={fill=white, at={(0.5,1.05,1.15)}, anchor=south, font=\scriptsize},
    reverse legend=true,
    xticklabels from table={graphs/data/energy_gpu.data}{Model},
    xticklabel style={rotate=45}
]

\addplot [fill=darkgoldenrod, postaction={pattern=crosshatch dots, pattern color=black, line width=0.4pt}]
table[
    x expr = \coordindex,
    y = acc_const,
    ]{graphs/data/energy_gpu.data};

\addplot [fill=asparagus, postaction={pattern=crosshatch dots, pattern color=black, line width=0.4pt}]
table[
    x expr = \coordindex,
    y = acc_static,
    ]{graphs/data/energy_gpu.data};

\addplot [fill=lightgray, postaction={pattern=crosshatch dots, pattern color=black, line width=0.4pt}]
table[
    x expr = \coordindex,
    y = acc_l2_dram,
    ]{graphs/data/energy_gpu.data};

\addplot [fill=lightblue, postaction={pattern=crosshatch dots, pattern color=black, line width=0.4pt}]
table[
    x expr = \coordindex,
    y = acc_reg_l1,
    ]{graphs/data/energy_gpu.data};

\addplot [fill=teal, postaction={pattern=crosshatch dots, pattern color=black, line width=0.4pt}]
table[
    x expr = \coordindex,
    y = acc_core,
    ]{graphs/data/energy_gpu.data};
    
\end{axis}

\begin{axis}[hide axis, bar shift=7.5pt,
    width=\columnwidth,, height=3.6cm,
    bar width=5pt,
    ybar stacked,
    ymin=0,
    ymax=1.5,
    xtick=data,
    legend columns=4,
    legend columns=8,
    legend style={fill=white, at={(0.5,0.9,1.15)}, anchor=south, font=\scriptsize},
    reverse legend=true,
    xticklabels from table={graphs/data/energy_gpu.data}{Model},
    xticklabel style={rotate=45}
]

\addplot [fill=darkgoldenrod, postaction={pattern=north east lines, pattern color=black, line width=0.4pt}]
table[
    x expr = \coordindex,
    y = bal_const,
    ]{graphs/data/energy_gpu.data};

\addplot [fill=asparagus, postaction={pattern=north east lines, pattern color=black, line width=0.4pt}]
table[
    x expr = \coordindex,
    y = bal_static,
    ]{graphs/data/energy_gpu.data};

\addplot [fill=lightgray, postaction={pattern=north east lines, pattern color=black, line width=0.4pt}]
table[
    x expr = \coordindex,
    y = bal_l2_dram,
    ]{graphs/data/energy_gpu.data};

\addplot [fill=lightblue, postaction={pattern=north east lines, pattern color=black, line width=0.4pt}]
table[
    x expr = \coordindex,
    y = bal_reg_l1,
    ]{graphs/data/energy_gpu.data};

\addplot [fill=teal, postaction={pattern=north east lines, pattern color=black, line width=0.4pt}]
table[
    x expr = \coordindex,
    y = bal_core,
    ]{graphs/data/energy_gpu.data};

\end{axis}

\begin{axis}[hide axis,
    bar shift=7.5pt,
    width=\columnwidth,, height=3.6cm,
    bar width=5pt,
    ybar stacked,
    ymin=0,
    ymax=1.5,
    xtick=data,
    legend columns=4,
    legend columns=8,
    legend style={fill=white, at={(0.5,0.9,1.15)}, anchor=south, font=\scriptsize},
    reverse legend=true,
    xticklabels from table={graphs/data/energy_gpu.data}{Model},
    xticklabel style={rotate=45}
]

\addplot [fill=darkgoldenrod]
table[
    x expr = \coordindex,
    y = extra,
    ]{graphs/data/energy_gpu.data};

\addplot [fill=asparagus]
table[
    x expr = \coordindex,
    y = extra,
    ]{graphs/data/energy_gpu.data};

\addplot [fill=lightgray]
table[
    x expr = \coordindex,
    y = extra,
    ]{graphs/data/energy_gpu.data};

\addplot [fill=lightblue]
table[
    x expr = \coordindex,
    y = extra,
    ]{graphs/data/energy_gpu.data};

\addplot [fill=teal]
table[
    x expr = \coordindex,
    y = extra,
    ]{graphs/data/energy_gpu.data};

\legend{Const, Static, L2, Reg+L1, Core}
\end{axis}

\begin{axis}[hide axis,
    bar shift=7.5pt,
    width=\columnwidth,, height=3.6cm,
    bar width=5pt,
    ybar stacked,
    ymin=0,
    ymax=1.5,
    xtick=data,
    legend columns=4,
    legend columns=8,
    legend style={fill=white, at={(0.5,-0.85,1.15)}, anchor=south, font=\scriptsize},
    reverse legend=false,
    xticklabels from table={graphs/data/energy_gpu.data}{Model},
    xticklabel style={rotate=45}
]

\addplot [fill=white, postaction={pattern=horizontal lines}]
table[
    x expr = 0,
    y = extra,
    ]{graphs/data/energy_gpu.data};
\addlegendentry{\em W8A8}

\addplot [fill=white, postaction={pattern=crosshatch, pattern color=black, line width=0.4pt}]
table[
    x expr = 0,
    y = extra,
    ]{graphs/data/energy_gpu.data};
\addlegendentry{\em HALO(perf-opt)}

\addplot [fill=white, postaction={pattern=crosshatch dots, pattern color=black, line width=0.4pt}]
table[
    x expr = 0,
    y = extra,
    ]{graphs/data/energy_gpu.data};
\addlegendentry{\em HALO(acc-opt)}

\addplot [fill=white, postaction={pattern=north east lines, pattern color=black, line width=0.4pt}]
table[
    x expr = 0,
    y = extra,
    ]{graphs/data/energy_gpu.data};
\addlegendentry{\em HALO(bal)}

\end{axis}

\end{tikzpicture}
	\caption{Normalized energy on GPUs.} 
	\label{fig:energy_gpu}
\end{figure}
Fig.~\ref{fig:energy_gpu} shows normalized energy consumption across constant, static, and dynamic sources. Constant power includes peripheral subsystems, while dynamic power comes from DRAM, caches, register files, and compute units. Although \textit{W8A8} uses the least energy due to uniformly low precision, its inability to adapt to workload sensitivity results in limited performance improvements. HALO\textit{(acc-opt)} improves execution time with a slight energy increase, while HALO\textit{(perf-opt)} and HALO\textit{(bal)} prioritize speed and still reduce energy on average, even though tiles run at higher VF points.

\textbf{HALO improves energy proportionality in the memory and core subsystems by targeting high-frequency execution only where needed, avoiding broad frequency increases that typically lead to energy spikes across the entire device.} This highlights HALO’s ability to balance performance and energy efficiency through hardware-aware optimization strategies tailored to specific deployment goals.
\section{6.\hspace{0.5cm}Conclusion}
\label{section:conclusion}
We present HALO, a timing-aware, weight-sensitive quantization framework that improves LLM inference efficiency. By selecting and quantizing weights based on circuit-level insights, HALO reduces power and critical-path-delays, enabling efficient DVFS while remaining hardware-compatible.

\section{Acknowledgments}
We thank the reviewers for their feedback and interesting discussion of this work.
This research is supported by the National Research Foundation, Singapore, under its Competitive Research Program Award NRF-CRP23-2019-0003 and the Ministry of Education, Singapore, under Tier 3 grant MOE-MOET32024-0003.

\bibliography{aaai2026}
\end{document}